\documentclass{elsart}

\usepackage{graphicx}
\usepackage{epsfig}
\usepackage{bm}

\usepackage{amssymb}

\begin{document}

\begin{frontmatter}





\title{Investigating the Geometrical Structure of \\ Disordered Sphere Packings }

\author{ T. Aste, M. Saadatfar, A. Sakellariou and T.J. Senden}
\address{ Department of Applied Mathematics, RSPhysSE, Australian National University, 0200 Australia.}


\address{\bf To appear on: { Physica A} (2004)}

\begin{abstract}
Bead packs of up to 150,000 mono-sized spheres with packing densities
ranging from 0.58 to 0.64 have been studied by means of X-ray Computed
Tomography.  These studies represent the largest and the most accurate
description of the structure of disordered packings at the grain-scale
ever attempted.  We investigate the geometrical structure of such
packings looking for signatures of disorder.  We discuss ways to
characterize and classify these systems and the implications that
local geometry can have on densification dynamics.
\end{abstract}


\end{frontmatter}


\section{Introduction}
\label{s.1}

For several centuries, the structures generated by tightly packed
mono-sized spheres have been studied by scientists as model systems for
understanding the emergence of order and crystallisation or -\textit{vice
versa}- the appearance of disordered and amorphous phases in natural
systems \cite{ppp}.  Despite all the efforts, several questions remain
unsolved.  
One of the most intriguing and challenging problems is to understand whether or not the notion of `ideal' or `typical' or `common' disordered packing configurations has any empirical basis \cite{Trusk00}.  
Indeed, disorder does not exclude organization and the positions of the packed spheres are locally highly correlated.
Experimentally, it is easy to observe that when spheres with
approximately equal sizes are poured into a jar, squeezed into a bag,
shaken or mixed, they form structures with densities in a narrow
range.  No stable packing configurations have been observed with
densities below 0.56 and no disordered assemblies of spheres have been
observed with densities above 0.645 \cite{ppp}.  Molecular dynamics
simulations show that hard sphere systems behave like a gas below
$\rho \sim 0.49$.  Upon compaction, liquid-like behaviour is observed
up to $\rho \sim 0.55$, where crystallisation starts to occur.  If
crystallization is avoided, the system undertakes a `glass' transition at $\rho \sim 0.56$ and then it can be compacted up to $\rho \sim 0.645$ where
no further densification can be induced \cite{Speed94b}.  
Empirical and simulated evidence suggest that something very special might
happen in the geometry of the packing at densities above $\rho \sim
0.56$; a process which must terminate below $\rho \sim 0.65$.  What
makes the understanding of this process particularly challenging is
that there are no \textit{a priori} reasons for the densification
process to stop around $\rho \sim 0.64$.  On the contrary, there are
plenty of local configurations which are denser than this limit and
any arrangement of spheres in stacked planar hexagonal closed packed layers (the so called Barlow packings \cite{Conway96}) can reach the density of $\rho = \pi/\sqrt{18} \sim 0.74$, as achieved in \textit{fcc} (face-centered cubic) or \textit{hcp} (hexagonal closed-packed) crystalline packings.

\begin{figure}
\mbox{\epsfig{file=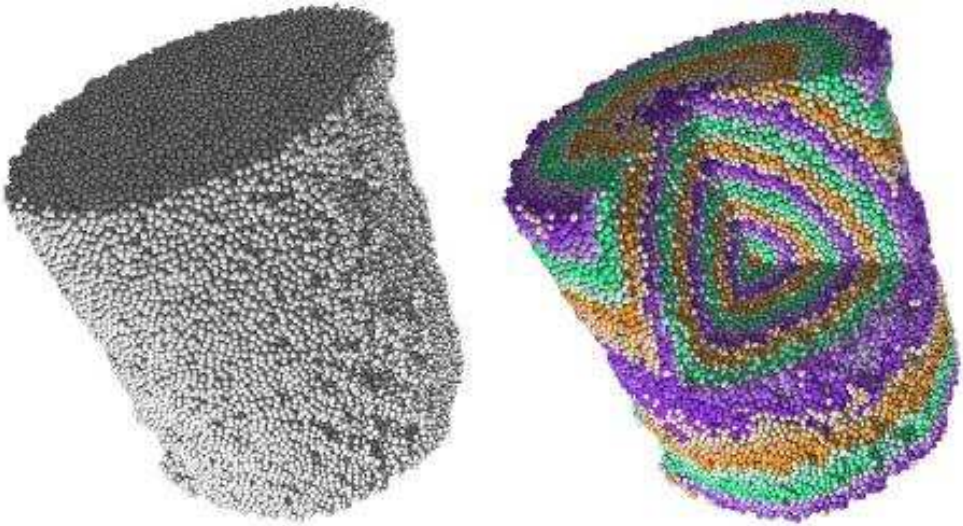,width=13.cm,angle=0}}
\caption{ (left) Volume rendering of $\sim 150,000$ sphere-pack in a
cylindrical container.  (right) Same image with the topological
distances from a given central sphere highlighted in colours (online
version).  }
\label{f.pack}
\end{figure}

\section{Experimental Apparatus and Methodology}
\label{s.2}
 
In this paper we report the analysis on 6 bead pack samples, all in a cylindrical container with an inner diameter of $D \simeq 55\; mm$ and filled to a height of $H \simeq 75\; mm$. 
Specifically:
\begin{itemize}
\item Two large samples with 150,000 beads and diameters $d =
1.00 \pm 0.05 \; mm$;
\item Four smaller samples with 35,000 beads and diameters $d = 1.59
  \pm 0.05 \; mm$.
\end{itemize}

A packing realization is shown in Fig.\ref{f.pack}.
To our knowledge, the present study is the first empirical investigation of \emph{large} packings of \emph{monosized} spheres.
Indeed, there are only two previous investigations which use a similar approach \cite{Seidler00,Richard03}, but the first concerns a single test-sample with $\sim 2,000$ spheres only, whereas the second investigates polydisperse assemblies with $\sim 15,000$ spheres.

{\bf Sample Preparation:}
Low density packings are obtained by placing a stick in the middle of the container and pouring the beads into it. 
Then, the stick is removed leaving a packing density $\rho \sim 0.58$ \cite{ppp}.  
Higher densities, up to $\rho \sim 0.63$, are achieved by gently tapping the sample.  
The densest sample at $\rho = 0.64$ is obtained by a combined action of gently tapping and compression from above (with the upper surface left unconfined at the end of the preparation). 
To reduce boundary effects, the inside of the cylinder have been made rough by randomly gluing spheres on the internal surfaces.

{\bf XCT Imaging:} A X-ray Computed Tomography apparatus (see
Sakellariou \textit{et al}. in this Issue \cite{Sakellariou04}) is
used to image the samples.  The two large samples were analysed by
acquiring data sets of $2000^3$ voxels with a spatial resolution of
$0.05 \;mm$; whereas the four small samples were analysed by acquiring
data sets of $1000^3$ voxels with a spatial resolution of $60 \;\mu m$.
After segmentation (see Sheppard \textit{et al} in this issue
\cite{Sheppard04}) the sample data sets are reduced to
three-dimensional binary images, representing two distinct phases, one
associated with the spheres and the other with air space.

{\bf Sphere Centres:} In order to proceed with the analysis of the
geometrical and statistical properties the position of all sphere centres are calculated from the binary images.  
Our approach is to find the sphere centres by moving a
reference sphere ($S$) throughout the binarised sphere pack ($P$) and
measuring the local overlap between $S$ and $P$. This corresponds to a
3-dimensional convolution: $P*S$. 
This method is made highly efficient by applying the convolution theorem which allow to transform the convolution into a product in Fourier space: ${\mathcal F}[P*S]= {\mathcal F}[P]{\mathcal F}[S]$, where ${\mathcal F}$ represents the (fast)Fourier Transform. 
The algorithm proceeds in 4 steps: {\bf 1)} fast Fourier transform of the
binary image (${\mathcal F}[P]$); {\bf 2)} transform the digitised map
of the reference sphere (${\mathcal F}[S]$, chosen with a diameter
about 10 \% smaller than $d$); {\bf 3)} perform the direct product
between these two; {\bf 4)} inverse-transform of the product:
${\mathcal F}^{-1}[{\mathcal F}[P]{\mathcal F}[S]]= P*S $.  
The result is an intensity map of the overlapping between the reference sphere and the bead pack, where the voxels closer to the sphere centres have a higher
intensity.  
A threshold on the intensity map, locates the groups of voxels surrounding the sphere centres. 
The centre of mass of these grouped voxels is a very good estimation
of the sphere centres in the pack.

{\bf Central Region:}
All the analyses reported hereafter have been performed over a central region ($\mathbf G$) at 4 sphere-diameters away from the sample boundaries.
Note that spheres outside $\mathbf G$ are considered when computing the neighbouring environment of spheres in $\mathbf G$.
The two large samples have about $N_G \sim 80,000$ spheres in ${\mathbf G}$, whereas the four smaller have about $N_G \sim 20,000$.
In Table \ref{t.1} the number of spheres in this region ($N_G$) is reported for each sample.

\begin{table}
\begin{tabular}[c]{llllllllll}
\hline
$\rho$ & $\sigma$ & $N_G$ & $N_c^{(1)}$ & $N_c^{(2)}$ & $N_c^{(3)}$ &
$T_D$ & {\it dis} & \textit{fcc} & \textit{hcp}  \\
\hline
0.586  & 0.005 & 84,444 & 2.75 & {\bf 5.26 }& 6.22 & 2.2 $\pm$ 0.2 & 27\% & 2\% & 3\% \\
0.593* & 0.006 & 18,763 & 3.15 & {\bf 5.91 }& 6.61 & 2.2 $\pm$ 0.3 & 29\% & 2\% & 3\% \\
0.617  & 0.005 & 82,626 & 3.23 & {\bf 6.11 }& 7.05 & 2.2 $\pm$ 0.1 & 34\% & 3\% & 7\% \\
0.626* & 0.008 & 19,522 & 3.56 & {\bf 6.18} & 7.25 & 2.5 $\pm$ 0.5 & 34\% & 3\% & 6\% \\
0.630* & 0.01  & 19,843 & 3.58 & {\bf 6.51 }& 7.40 & 2.5 $\pm$ 0.5 & 35\% & 3\% & 8\% \\
0.640* & 0.005 & 20,188 & 3.733& {\bf 6.94} & 7.69 & 2.7 $\pm$ 0.3 & 37\% & 4\% & 12\% \\
\hline
\end{tabular} 
\caption{ \label{t.1} Sample density ($\rho$), density fluctuations
($\sigma$), number of spheres in the central region ($N_G$), average
number of neighbours ($N_c$), topological density ($T_D$), percent of
local configurations with a given local order: disordered ({\it dis}),
\textit{fcc} - like and \textit{hcp} -like.  The symbols `*' in the first column indicate measures on the smaller samples ($\sim$~35,000 beds).  
The three quantities $N_c^{(1)}$, $N_c^{(2)}$ and $N_c^{(3)}$, correspond respectively to the average number of neighbours within the three radial distances: (1) $d$, (2) $d\! + \!v/2$ and (3) $ d\! + \!v$, where $v$ is the voxel-size.  }
\end{table}

\section{Study of the Packing Configurations and Discussion}
\label{s.3}

{\bf Densities:} We calculate the \emph{local-densities} and the
\emph{sample-densities}.  The local-densities are the fractions
between the sphere-volumes and the volumes of the Vorono\"{\i} cells
constructed around the centre of each sphere in the sample (the
Vorono\"{\i} cell is the portion of space closest to a given centre
in respect of any other centre).  
The sample-densities are fractions between the sum over the volumes of the spheres in $\mathbf G$ and the sum over the volumes of the Vorono\"{\i} cells associated with these spheres. 
We observe bell-shaped distributions with average densities in the range
$0.586 \le \rho \le 0.640$ and standard deviations $\sigma$ within 1.5
\% (see Table \ref{t.1}).  
However, the local densities are not homogeneously distributed in the sample.  Typically, the  densities are relatively smaller than the average in a region close to the cylinders central axis; the density increases going out from the centre then it saturates to rather homogeneous values up to a distance of a few (2-3) sphere diameters from the boundary.  
Rather inhomogeneous densities are also observed in the
vertical direction, but in this case we find different behaviours
depending on the sample-preparation.  A detailed analysis of these
behaviours will be the subject of a future publication.

{\bf Local Environment:} The identification of touching spheres
is, in general, an ill-defined problem from a experimental point of
view.  Indeed, the result is affected by the precision of the
calculated sphere centres and their diameters.  
At the present we are developing a new technique to overcome this problem.  
With the present data, we assume that the spheres in contact are located at a radial distance between the diameter $d$ and $d+v$, where $v$ is the
voxel-size.  
Table \ref{t.1} reports the values of the average number
of contacting neighbours ($N_c$) computed in $\mathbf G$ at the three different radial distances: $d$, $d+v/2$ and $d+v$.  
We observe values between $N_c \sim$ $3$ and $8$, and an increasing trend with the packing density.

\begin{figure}
\begin{center}
\begin{tabular}{ll}
\mbox{\epsfig{file=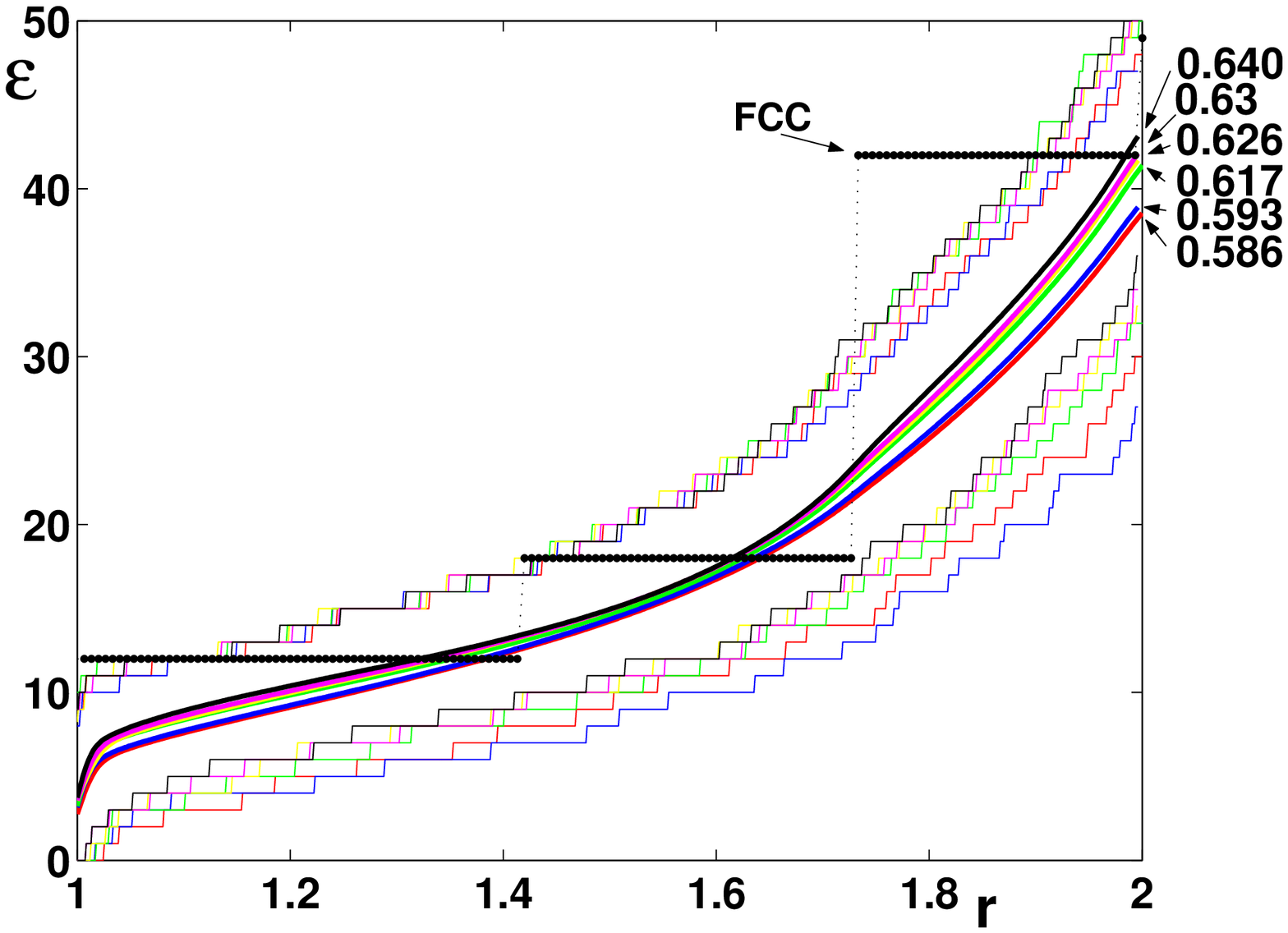,height=5.cm,angle=0}}
&\mbox{\epsfig{file=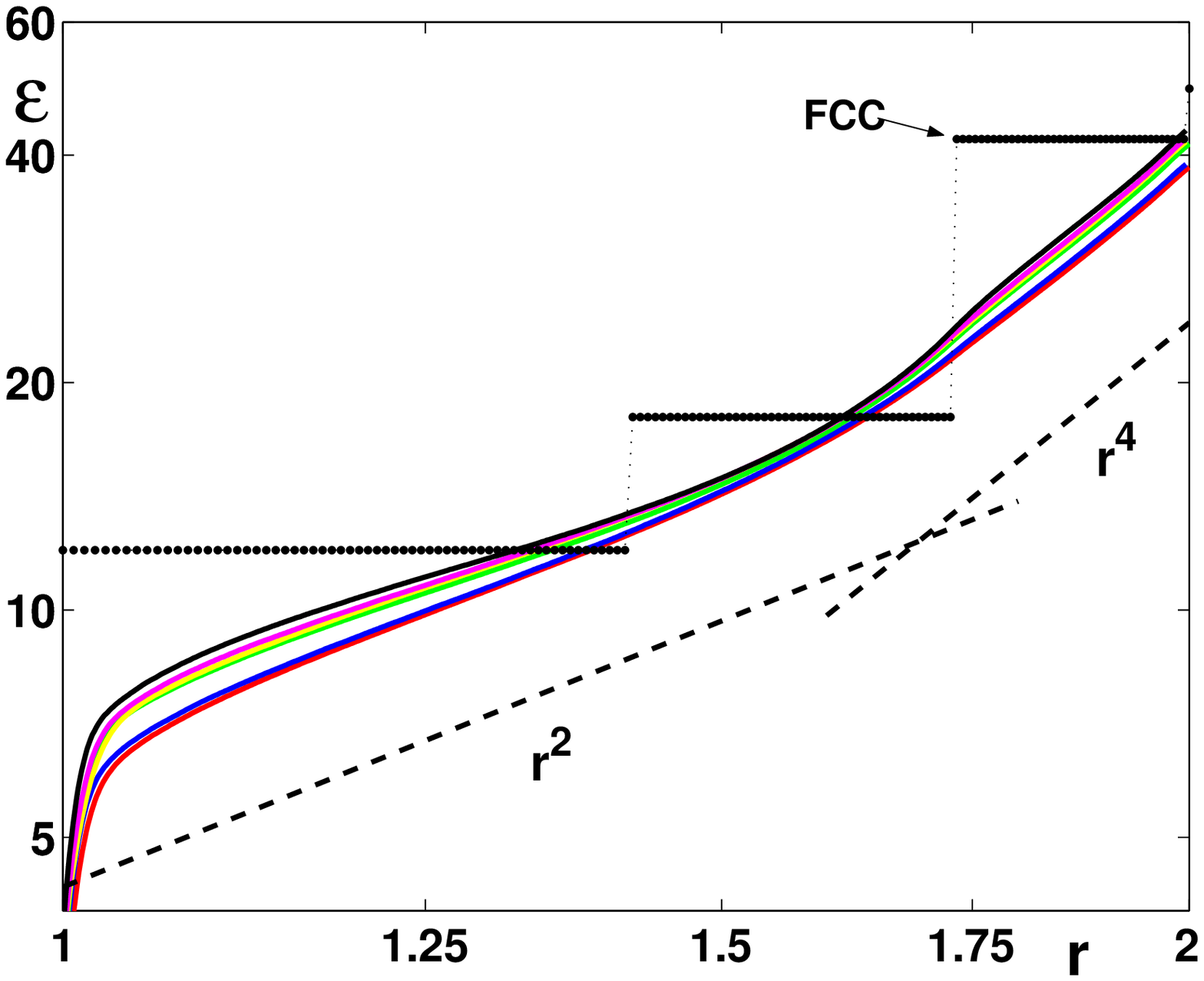,height=5.cm,angle=0}}
\end{tabular}
\end{center}
\caption{ Packing Efficiency $\epsilon$ v.s. radial distance (in
bead diameters).  (left) The thick lines refer to the average value within
a central region; the thin lines above and below
represent upper and lower limits for each sample; the dotted line is
the \textit{fcc} configuration. The colours (online only) are associated with the
different samples.  (right) The same plot of only the average values in
log-log scale; the two straight lines are references representing
respectively the power laws: $\epsilon \propto r^2$ and $\epsilon
\propto r^4$.}
\label{f.nn}
\end{figure}

{\bf Packing Efficiency:} We compute the number of spheres placed within a certain radial distance from a given sphere.  
This is a measure of how efficiently local dense agglomerate of spheres are formed and therefore we called it \emph{packing efficiency} $\epsilon$.  
It  is well known that no more than 12 spheres can be found in contact with one sphere (the `Kissing number' \cite{ppp}), but the upper limit for the number of spheres within a given radial distance is, in general, unknown.  
Here, we empirically investigate the local environment surrounding each sphere starting from the spheres in contact up to the ones at a radial distance of two bead diameters.  
Fig.\ref{f.nn} shows the average, the  maximum and the minimum numbers of neighbours within a given radial distance from any sphere in $\mathbf G$.
Interestingly, we observe that, in the regions around $r \sim 1.4$ and $1.7$, disordered packings show better average packing efficiency than \textit{fcc}.  Moreover, we observe that in all the samples the upper limits are above or equal to the \textit{fcc} packing-efficiency in a large range of the radial distances.  
Let us stress that, these surprisingly high efficiencies in disordered
packings could be the inner driving cause for the occurence of such
disordered assemblies.

\begin{figure}
\begin{center}
\begin{tabular}{ll}
\mbox{\epsfig{file=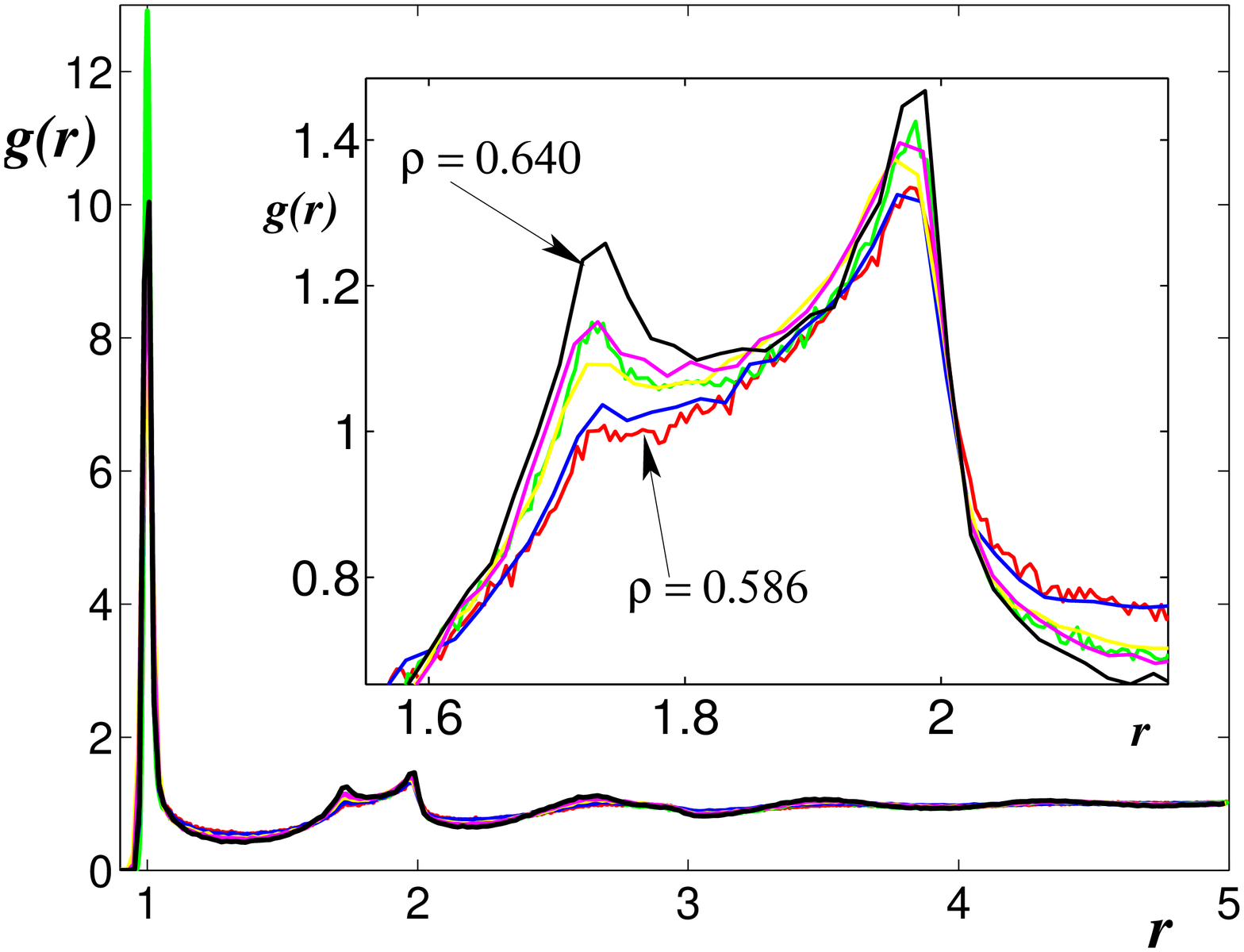,height=4.5cm,angle=0}}
&\mbox{\epsfig{file=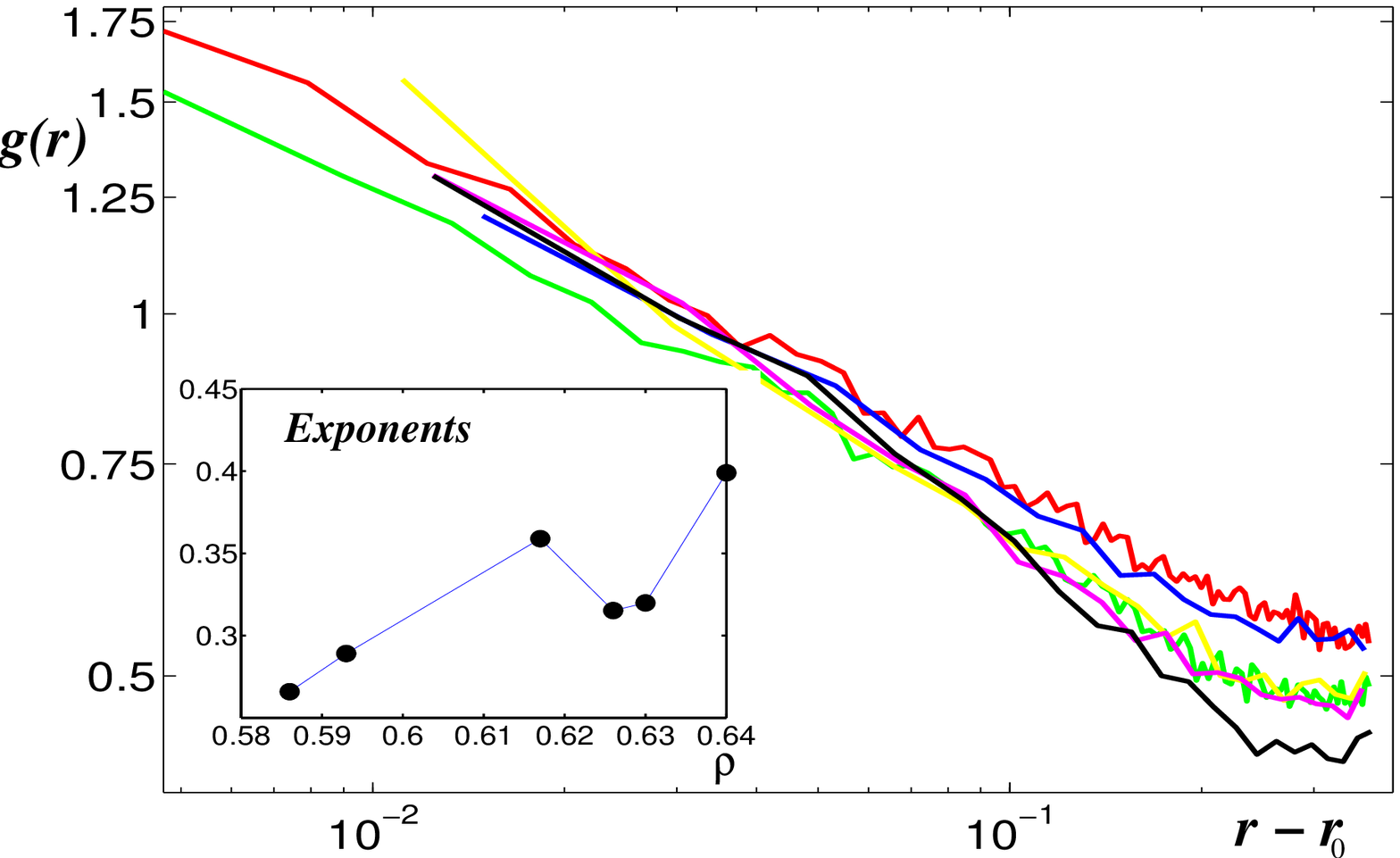,height=4.5cm,angle=0}}
\end{tabular}
\end{center}
\caption{ (left) Radial distribution function $g(r)$ v.s. radial distance $r$ (in
bead diameters) for all the samples examined.  
The insert shows the two peaks at $r = \sqrt{3}$ and $r = 2$ in detail.  
(right) In the region between $1 < r < 1.4$, the radial distribution function decreases as a power law: $g(r)\propto (r-r_0)^{-\alpha}$ (linear trend).  
The insert shows the best-fit estimations for the exponent $\alpha$, demonstrating an increasing trend with the density.
\label{f.gr}
}
\end{figure}

{\bf Radial Distribution Function:} The radial distribution function
$g(r)$ (shown in Fig.\ref{f.gr}) is calculated as the average number of sphere centres, within a radial distance $r - \Delta /2$ and $r + \Delta /2$, divided by $c r^2$.  
The constant $c$ is fixed by imposing that asymptotically $g(r)\rightarrow 1$ for $r \rightarrow \infty$.  
(We have verified that different choices of $\Delta$ within a broad range of $10^{-4}$ to $10^{-2}$ (in bead diameters units), lead to almost indistinguishable results.)
The detail of the close-neighbour region (Fig.\ref{f.gr}) shows that the two peaks at $r=\sqrt{3} $ and $r = 2$ (in bead diameters) both increase in height with the packing density.
This is an indication that there is an increasing number of configurations in the contact network \cite{ContactNetwork} with edge-sharing in-plane triangles (peak at $\sqrt{3} $) and with three (or more) aligned spheres (peak at $ 2 $).  
This might indicate an increasing organisation in the packing structure but, on the other hand, no clear signs of crystallisation were detected (see below).

For all the samples investigated, we found that in the range $r_0 \le
r \le 1.4$, the radial distributions follow a power law behaviour:
$g(r) \propto (r-r_0)^{-\alpha}$, with exponents $\alpha$ between 0.3
and 0.45 and the singularity at $r_0 = 1.03$ (Fig.\ref{f.gr}).
A similar behaviour, but with $\alpha = 0.5$ and $r_0 = 1$, was reported in \cite{Silb02} for molecular dynamics
simulations.  In Fig.\ref{f.gr} it is also highlighted the growing
trend of the exponent $\alpha$ with the density $\rho$.  
To our knowledge, this is the first time that such behaviours (still theoretically unexplained) are observed in experiments.

{\bf Shell Analysis:} The sphere neighbourhood environment has been also studied topologically by performing a \emph{shell analysis} on the contact network and calculating the topological density ($T_D$) \cite{Brunner71,Brunner79,OKeffe91,OKeffe95,Aste99shell}.  
We observe that the number of spheres at a given topological distance ($j$ \cite{noteTD}) from a central one follows quite accurately the quadratic law:
$K_j = 3 T_D j^2 + c_1 j + c_0$.  
The best fit values of $T_D$'s are reported in Table \ref{t.1}. 
(This analysis was performed by defining in contact spheres at radial distance within $r \le d+v/2$; the dependence of $T_D$ on the choice of the contact-neighbour criteria will be discussed in a forthcoming paper). 
We note that the topological densities increase with the sample-densities and are consistently smaller than $10/3 = 3.33...$, which is the theoretical lower limit for Barlow packings \cite{Conway96}.  
A view of the topological shell structure is shown in Fig.\ref{f.pack}.

{\bf Local Order:} We estimate the local packing orientation by associating a set of spherical harmonics to the vectors $\vec r$ between a sphere and the set of neighbouring spheres ($|\vec r| \le 1.2$ - in beads diameters): $Y_{l,m}(\theta(\vec r),\phi(\vec r))$ (with $\theta(\vec r)$ and $\phi(\vec r)$ the polar and azimuthal angles associated with $\vec r$).
Following \cite{Steinh83}, a set of rotationally invariant quantities are calculated:
\begin{equation}
Q_l = \sqrt{ \frac{4 \pi}{2l+1} \sum^l_{m=-l} 
|\sum_{i \in \mathbf G} Y_{l,m}(\theta(\vec r_i),\phi(\vec r_i))|^2}\;\;,
\label{ee}
\end{equation}
We observe that the $Q_l$'s (with $l =4,6$) calculated over many sub-volumes of the samples tend to vanish as the sub-volumes become larger.  
This indicates that these packings are macroscopically like \emph{isotropic liquids} \cite{Steinh83} and across the entire system there is no prevailing orientational order.  
On the other hand, even if \emph{globally} there are no signs of order or
organisation, this method is useful to investigate the orientational
properties of the \emph{local} configurations.  In particular, for
each sphere in the central region we measured the coefficients $Q_4$
and $Q_6$ calculated from the angles between a sphere and all its
neighbour up to a radial distance $r=1.2$.  We first search for
signatures of local crystalline orientational order measuring the
fraction of local configurations with $(Q_4,Q_6)$ in a region close to
the \textit{fcc} $(0.1909,0.5745)^{(fcc)}$ and \textit{hcp}
$(0.0972,0.4848)^{(hcp)}$ order.  We observe that a relatively small
fraction of configurations have $Q_4$ and $Q_6$ within a range $\pm
0.05$ of their ideal \textit{fcc} and \textit{hcp} values (see `\textit{fcc}' and `\textit{hcp}' in Table \ref{t.1}).  
We also verify that other special configurations like \textit{Icosahedral}, \textit{bcc} and \textit{Simple Cubic}, have significantly lower occurrences.  
On the other hand, we observe that large percentages (between 27\% and 37\%) of the local configurations share a common signature in their orientational order with $0.15 \le Q_4 \le 0.25$ and $0.4 \le Q_6 \le 0.5$ (see `{\it dis}' in Table \ref{t.1}).  
To understand whether this is an indication of a `typical' local disordered packing around $(0.20,0.45)^{(dis)}$ is the matter of present investigations.

{\bf Structural Arrest:} We study a quantity which is relevant for system dynamics: the \emph{escape probability} which is the probability that a sphere can moves outward from a given local configuration without re-adjusting the positions of its first neighbours \cite{AsCo03}.  
This quantity is calculated by constructing circles through the centres of the three spheres corresponding to the three faces incident at each vertex in the Vorono\"{\i} polyhedron.  
If one of these circles has a radius larger than $d$, it implies that the
central sphere can pass through that neighbouring configuration and
move outward from its local position without displacing the
first-neighbours.  
In other words, the neighbouring `cage' is \emph{open} if at least one radius is lager than $d$; \textit{vice versa} the cage is \emph{closed} when all radii are smaller than $d$.  
The escape probability is defined as the fraction of open cages.
We find that all the samples with $\rho > 0.6$ have zero escape probability in $\mathbf G$.
Whereas, the two samples with $\rho = 0.586$ and $\rho = 0.595$, have very small fractions of open cages (0.1\% and 0.6\%, respectively).

Following a theoretical approach recently developed by one of the
authors \cite{AsCo03}, the packing realization can be considered as
an inherent structure and the (thermo)dynamical approach toward this configuration can be reconstructed by virtually decreasing the sphere-diameters reducing the effective density.  (Here, the main underlying assumption is that the
system dynamics before the structural arrest develops through
configurations around the final inherent-structure realization.)
We find that, for all the samples with $\rho > 0.6$, the escape
probabilities go to zero at effective densities between 0.61 and 0.63.  
This strongly suggests that around $\rho \sim 0.62 \pm 0.01$ an important phase in the system dynamics reaches an end: above this density, local readjustments involving only the displacement of a single sphere are forbidden and the system compaction can proceed only by involving the collective and correlated readjustment of larger set of spheres.

{\bf Acknowledgements}
The authors thank Ajay Limaye for the preparations of Fig.\ref{f.pack}.
Senden gratefully acknowledges the ARC for his Fellowship.
This work was partially supported by the ARC discovery project DP0450292.





\begin{thebibliography}{}


\bibitem{ppp}
T. Aste and D. Weaire 
``The Pursuit of Perfect Packing'' (Institute of Physics, Bristol 2000).

\bibitem{Trusk00}
T.M. Truskett, S. Torquato and P.G. Debnedetti 
``Towards a quantification of disorder in materials Distinguishing equilibrium and glassy sphere packings''
{\it Phys. Rev. E} {\bf 62} (2000) 993.

\bibitem{Speed94b}
R. J. Speedy 
``On the reproducibility of glasses'', 
{\it J. Chem. Phys.} {\bf 100} (1994) 6684.


\bibitem{Conway96}
J.H. Conway, , N.J.A. Sloane,
``Low Dimensional Lattices VII: Coordination Sequences''
{\it  Proc. Royal Soc. London A} {\bf 453} (1997) 2369.

\bibitem{Seidler00}
G. T. Seidler, G. Martinez, L. H. Seeley, K. H. Kim, E. A. Behne, S. Zaranek,
B. D. Chapman, S. M. Heald, 
``Granule-by-granule reconstruction of a sandpile from x-ray microtomography data'', 
{\it Phys. Rev. E} {\bf 62} (2000) 8175.

\bibitem{Richard03}
P. Richard, P. Philippe, F. Barbe, S. Bourl\`es, X. Thibault, and D. Bideau 
``Analysis by x-ray microtomography of a granular packing undergoing compaction'' 
{\it Phys. Rev. E.} {\bf 68} (2003) 020301.

\bibitem{Sakellariou04}
A. Sakellariou, T. J. Sawkins, T. J. Senden and A. Limaye,
``X-ray Tomography for Mesoscale Physics Applications''
{\it Physica A}, In this Issue, (2004) p....

\bibitem{Sheppard04}
A.P. Sheppard, R.M. Sok and H. Averdunk,
``Techniques for Image Enhancement and Segmentation of Tomographic Images of Porous Materials'',
{\it Physica A}, In this Issue, (2004) p....


\bibitem{ContactNetwork}
The contact-network is the network formed by linking with edges the centers of spheres in contact.


\bibitem{Silb02}
L. E. Silbert, D. Ertas, G. S. Grest, T. C. Halsey and D. Levine,
``Geometry of frictionless and frictional sphere packings''
{\it Phys. Rev. E} {\bf 65} (2002) 031304.



\bibitem{Brunner71}
G. O. Brunner and F. Laves,
``Zum Problem der Koordinationszahl''
{\it Wiss. Z. Techn. Univ. Dresden} {\bf 20} (1971) 387-390.

\bibitem{Brunner79}
G. O. Brunner,
``The properties of coordination sequence and conclusions regarding the lowest possible density of zeolites''
{\it J. Solid State Chem.} {\bf 29} (1979) 41-45.

\bibitem{OKeffe91}
M. O'Keeffe, 
``N-dimensional diamond, sodalite and rare sphere packings''
{\it Acta Cryst.} {\bf A 47} (1991) 748-753.

\bibitem{OKeffe95}
M. O'Keeffe, 
``Coordination sequences for lattices''
{\it Zeit. f. Krist. } {\bf 210} (1995) 905-908.


\bibitem{Aste99shell}
T. Aste, 
``The shell map'',
in Foams and Emulsions, eds. J. F. Sadoc and N.
Rivier, (Kluwer Academic Publisher, Netherlands 1999) 497-
510.


\bibitem{noteTD}
Starting from a given sphere (at topological distance $j=0$) its first neighbours are at distance $j=1$, then the neighbours of the neighbours (which have not been already assigned with a distance) are at $j=2$, etc..


\bibitem{Steinh83}
P. J. Steinhardt, D. R. Nelson and M. Ronchetti,
``Bond-orientational order in liquids and glasses'',
{\it Phys. Rev. B} {\bf 28} (1983) 784.

\bibitem{AsCo03}
T. Aste and A. Coniglio,
``Cell approach to glass transition''
{\it J. Phys.: Condens. Matter} {\bf 15} (2003) S803.


\end{thebibliography}
\end{document}